\documentclass[runningheads, envcountsame, a4paper]{llncs}
\usepackage{amssymb}
\usepackage{mathtools}
\usepackage{dsfont,physics}
\usepackage{mathrsfs}
\usepackage[pdftex]{graphicx} 
\usepackage[hyphens]{url}
\usepackage{epstopdf} 
\usepackage{microtype} 
\usepackage[misc,geometry]{ifsym} 
\usepackage{dcolumn}
\usepackage{verbatim,comment} 
\usepackage{float}
\usepackage{color}
\usepackage[utf8]{inputenc}
\usepackage{appendix}
\usepackage{diagbox}
\usepackage{multirow,hhline}
\usepackage[table]{xcolor}
\usepackage{cleveref}
\usepackage{makecell}

\begin{document}

\title{Quantum Candies and Quantum Cryptography}

\author{Junan Lin \inst{1}\orcidID{0000-0002-3096-931X}\thanks{Corresponding author.} \and
Tal Mor\inst{2}}
%
%
\institute{Institute for Quantum Computing and Department of Physics and Astronomy, University of Waterloo, Waterloo, Ontario N2L 3G1, Canada\\
\email{j242lin@uwaterloo.ca}
\and
Computer Science Department, Technion – Israel Institute of Technology, Technion city, Haifa 3200003, Israel\\
\email{talmo@cs.technion.ac.il}}
\maketitle
\setcounter{footnote}{0} 

\begin{abstract}
The field of quantum information is becoming more known to the general public.
However, effectively demonstrating the concepts underneath quantum science and technology to the general public can be a challenging job.
We investigate, extend, and much expand here ``quantum candies'' (invented by Jacobs), a pedagogical model for intuitively describing some basic concepts in quantum information, including quantum bits, complementarity, the no-cloning principle, and entanglement.
Following Jacob's quantum candies description of the well known quantum key distribution protocol BB84, we explicitly demonstrate various additional quantum cryptography protocols using quantum candies in an approachable manner.
The model we investigate can be a valuable tool for science and engineering educators who would like to help the general public to gain more insights about quantum science and technology: most parts of this paper, including many protocols for quantum cryptography, are expected to be easily understandable by a layperson without any previous knowledge of mathematics, physics, or cryptography.

\end{abstract}

\section{Introduction}

Quantum information science and technology is a growing field that provides interesting concepts such as quantum computers, quantum teleportation, and quantum cryptography.
Its development has proved fruitful in both theoretical and experimental aspects, leading to various demonstrations in research labs. More recently the hi-tech became majorly involved, with companies such as IBM, Google, Intel, Microsoft, and Alibaba (plus many  startup companies)  investing in quantum devices for computing and/or communication and cryptography. 

The relevant ``quantum concepts'', in their simplest forms, can be demonstrated with simple  information units known as quantum binary digits (a binary digit gets one of two values, zero or one), 
named quantum bits or qubits.
Many applications of quantum cryptography, including quantum key distribution (QKD) protocols~\cite{bennett1984quantum,bennett1992quantum,Ekert1991quantum},  can be explained using qubits.
While the basic ideas behind quantum information in general, and especially behind these QKD protocols, are not very complicated for an ``insider'', they can be abstruse for someone without the necessary physics or mathematics background.
This hinders general audiences from correctly understanding these concepts.
An intuitive description of such concepts will potentially be beneficial not only for the general public, but also for researchers in other fields to understand and visualize protocols. 

In this work, we follow --- and much expand --- a simple but powerful model of ``quantum candies'' (or in short, ``qandies'') originally suggested by Jacobs~\cite{Kayla2009}, who proposed a new type of very special candies plus weird machines that produce those special candies. The non-conventional properties of the quantum candies and those machines --- plus our wide and wild expansions --- are used here as  explanatory tools for various quantum concepts and technologies. 

The model had originally been developed in two (independent) steps: Karl Svozil presented ``chocolate balls'' in several papers (e.g.~\cite{svozil2006staging}), including using these on an actual stage to demonstrate a pseudo-quantum model that resembles quantum cryptography. 
Kayla Jacobs, independently~\cite{Kayla2009}, invented an intuitive and much-closer-to-quantum variant based on properties of hypothetical candies having two different colors and two different tastes. 
To the best of our knowledge, Jacobs' model was presented in seminars (at MIT and at the Technion, and mainly to high-school students) but has never been officially published.

The paper is structured as follows.
In \cref{candy section} we review Svozil's and Jacobs' demonstrations of the most well-known QKD protocol: the BB84 protocol~\cite{bennett1984quantum}. 
We then propose several straightforward and more sophisticated extensions (including the famous B92 protocol~\cite{bennett1992quantum} and more).
In \cref{Sec:qandies-concepts} we expand the quantum candies model to explain various quantum concepts including entanglement, and an attack on ``quantum bit commitment'' in \cref{Sec:bit-commitment}.
We end with some brief closing remarks in \cref{Sec:Discussion}.

\section{Quantum Cryptography, Chocolate Balls, Quantum Candies, the BB84 Protocol, and Beyond}\label{candy section}
Suppose that two users, Alice and Bob, would like to communicate secretly. They may do so if they share a secret random and sufficiently long key (typically, a string of bits). Thus, the remaining problem is how to distribute an identical random and secure key.
Quantum key distribution (QKD) protocols solve the key distribution problem by utilizing quantum objects that, due to the rules of quantum theory, cannot be copied, and any attempt to copy them or even slightly obtain information from them enables Alice and Bob to detect the presence of the eavesdropper.
Thus an eavesdropper can block the communication but cannot learn the secrets. 

The BB84 QKD protocol proposed by Bennett and Brassard in 1984~\cite{bennett1984quantum}, is simple and it also can serve as an excellent tool for understanding the basics of quantum theory. Svozil and Jacobs found simple \emph{and sweet} ways to describe the BB84 protocol in somewhat-classical ways.

\subsection{Chocolate Balls and Generalized Urn Model}\label{GU section}

Svozil~\cite{svozil2006staging} described a simple way to stage the BB84 protocol using real-life chocolate balls, based on the so-called generalized urn (GU) model by Wright~\cite{wright1990generalized}.
In this model, the qubits are represented by chocolate balls wrapped in black foil with two binary numbers printed on the surface, one binary digit (0 or 1) in red color and the other binary digit (0 or 1) in green.
If we denote the case of a zero written in red and a zero written in green as
\{0,0\}[red,green], then the four types of balls include also 
\{0,1\}[red,green], \{1,0\}[red,green] and \{1,1\}[red,green].

Any user can view the numbers on the chocolate balls, but 
\emph{must obey} the restriction that he/she must put on red-filtering or green-filtering glasses first, so that only one binary number can be seen every time a person looks at a ball. 

The chocolate balls are drawn from a large urn containing an equal number of the four types of balls.
To send a string of random bits to Bob, Alice wears a pair of colored glasses, randomly draws a single ball from the urn, and records the number she sees so she keeps the data regarding both the digit and its color. 
The ball is then sent to Bob who, before obtaining it, randomly picks a glasses color and wears the glasses. 
Only then Bob is allowed to observe it, and then to record the color and number (as Alice did).

After repeating these steps as many times as they wish, each time again randomly picking a new pair of colored glasses, Alice and Bob then announce their choices of colors publicly, and keep the bits obtained only in the cases where they picked the same color of glasses, which guarantees that they will obtain the same bit string. 
About half the data is thrown away, in case all previous choices were made at random.

If the transmission between Alice and Bob for the chocolate balls is not eavesdropped, or if eavesdropped by an eavesdropper that \emph{follows the rules}, then the distributed key can be trusted.  
However, a cheating eavesdropper (we call her Eve) can obtain \emph{all} the transmitted information by simply not obeying the imposed law of wearing colored glasses.
Moreover, her observations do not change the chocolate balls.
Svozil~\cite{svozil2014non} analyzed various urn models and their connections to quantum physics, and of course was aware that this demonstration of BB84 only partially illustrates the BB84 protocol, as it does not illustrate its true quantumness or the resulting security.




\subsection{Jacobs' Quantum Candies Model}\label{BB84 section}
We now consider Jacobs' model, where, instead of chocolate balls, there exists candies to which we gradually assign ``quantumness''.
We call the resulting sweets ``quantum candy'' (or, \emph{qandy},
or Jacob's qandies). The model was originally presented by Kayla  Jacobs~\cite{Kayla2009} as an easy way to present quantum bits and QKD to the general audience.

Jacobs considered candies that have two different \emph{general properties}, color and taste. 
The color of a candy can be red (R) or green (G), while the taste can be chocolate (C) or vanilla (V). 

At first, the above general properties of color and taste and specific properties (red, green, chocolate, vanilla) show much similarity to Svozil's sweets, and may actually even found to be fully equivalent! Here are the four options of chocolate balls versus four options of candies, with an obvious one-to-one correspondence:
\begin{gather*}
    \{0,0\}\text{[red,green]} \longleftrightarrow \{C,R\}\text{[taste,color]}\\
    \{0,1\}\text{[red,green]} \longleftrightarrow \{C,G\}\text{[taste,color]}\\
    \{1,0\}\text{[red,green]} \longleftrightarrow \{V,R\}\text{[taste,color]}\\
    \{1,1\}\text{[red,green]} \longleftrightarrow \{V,G\}\text{[taste,color]}
\end{gather*}
It is trivial to see that this translation from the generalized urn model is exact thus far, hence if an eavesdropper (on BB84) could \emph{both} look and taste, the candies protocol becomes as insecure as the generalized urn model once Eve does not obey the rules. 

\begin{figure}[!ht]
	\centering
	\includegraphics[width=0.55\textwidth]{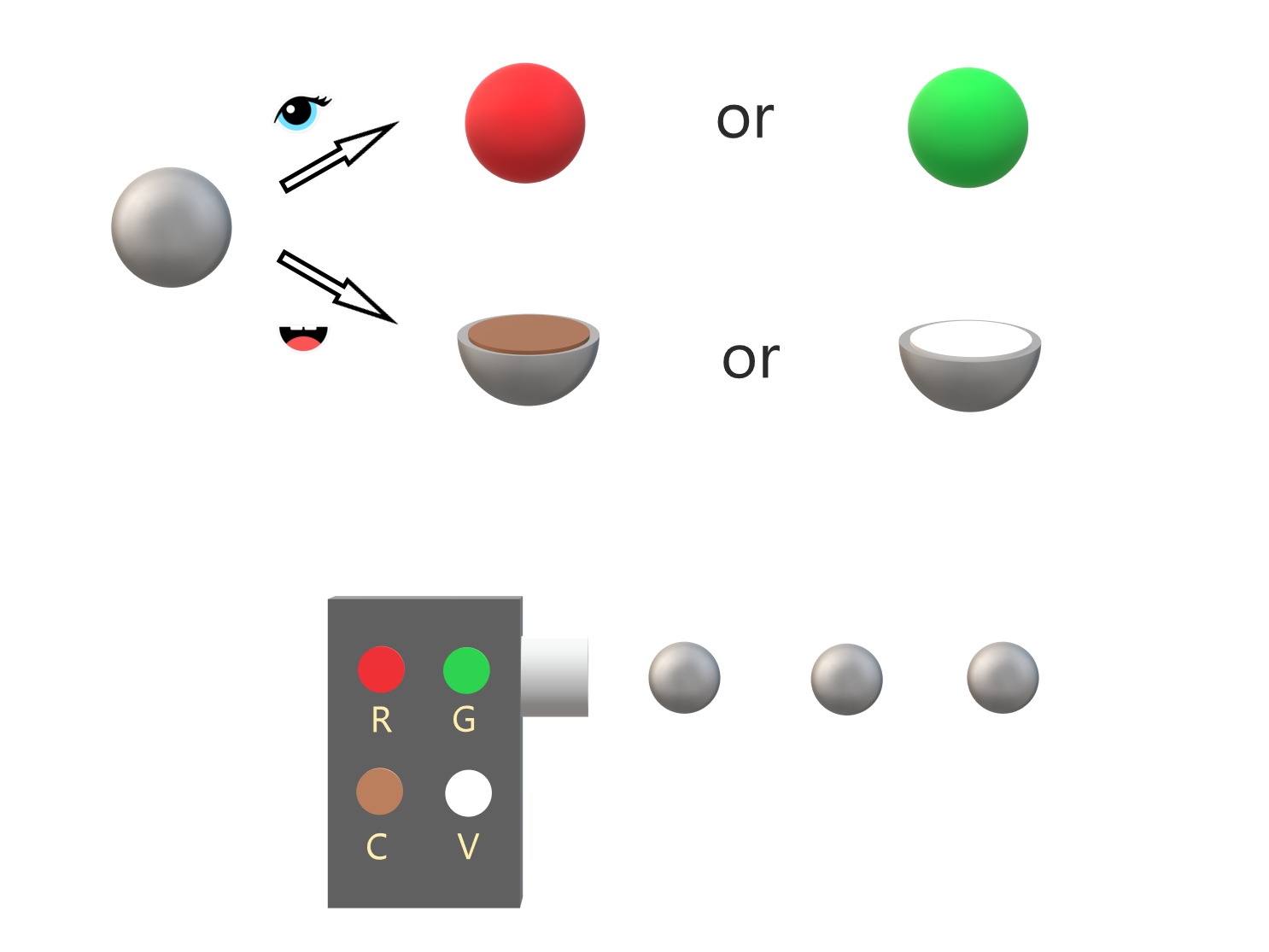}
	\caption{Top: illustrative figures of qandies with two general properties, color and taste.
	Before any choice of observation, all qandies appear identical. If one looks at a qandy, it will appear as red or green, but one cannot taste it again; if one tastes a qandy, he/she would taste either chocolate or vanilla, but one cannot then learn anything about its color.
	Bottom: a qandy-producing machine.
	The user can press one of the four buttons on the machine and generate the corresponding qandy. Only one general property can be fixed for every qandy produced, while the other general property would be randomly assigned by the machine (if we follow Einstein) or have the other property non-existing at all (if we follow modern understanding and Jacob's qandies).} 
	\label{candy-fig}
\end{figure}

As described in \cref{candy-fig}, we impose some unusual rules onto these candies that make them ``\emph{qandies}''. First, each qandy has only a single specific property:
\begin{equation*}
    \{C\}\text{[taste]},\ \{V\}\text{[taste]},\ \{R\}\text{[color]},\ \{G\}\text{[color]}
\end{equation*}
What are the implications from this rule in terms of generating the qandies and in terms of ``observing'' them?
First, a user can \emph{only} learn about one general property of a qandy, but not both. Namely, if one looks at a qandy, it  will  appear  as  red  or  green,  but  one  cannot  taste  it  anymore (its taste is destroyed by looking at it);  if  one  tastes  a  qandy, he/she would taste either chocolate or vanilla, but one cannot learn anything about its color anymore (its color is destroyed).
Second, a qandy-making machine has four buttons, one for choosing $C$, one for choosing $V$, one for choosing $R$ and one for choosing $G$.
Third, if the machine generates (as an example) a chocolate qandy ($C$), one can taste and know it is chocolate, but if one looks, a random color is seen ($R$ or $G$) and the taste is destroyed.

The key feature of these qandies is that each single qandy \textbf{really} has only a single specific property.
This is one form of what is known as the complementarity principle\footnote{Other forms of the principle exist, e.g., that elementary particles like electrons or photons have both particle characteristics and wave characteristics.} in quantum physics: if color is defined, taste cannot be defined, and if taste is defined color cannot be defined. The rule of complementarity applies both to the person (i.e. person/machine) preparing the qandies, and to anyone observing the qandy.

But maybe an extremely sophisticated (having some future technology) eavesdropper can somehow know both the color and taste?
Surprisingly, Einstein's view of quantum physics, e.g. his famous sentence ``God does not play dice'' and also the EPR paradox~\cite{EPR35}, is also in complete correspondence to the above.
In his view quantum physics is incomplete: it has ``hidden varaibles'' and a deeper theory (not yet known to Science, but maybe known to that futuristic Eve) may unveil the hidden variables and hence will provide both the color and the taste of these candies.   

If we try to use our ``classical imagination'' to \emph{build} the machine, we might think as Einstein did: imagine a futuristic machine which produces standard classical candies that are wrapped with an identical piece of edible but opaque candy paper, such that none of the candy's properties can be known by only looking at a freshly prepared candy.
The machine is a black box to all users, who can only interact with the machine through four external buttons: $\{C\},\{V\},\{R\},\{G\}$.
If Alice presses $\{R\}$ or $\{G\}$ to prepare a candy with definite color, that candy does not have a defined taste, and \emph{vice versa} --- when one general property is well defined, the other general property becomes \emph{random}. 
So any user cannot fix or learn both the color and taste of a candy.
And indeed, if Alice's produces a candy with a specific color, say $\{R\}$, and Bob tastes that candy, he will taste a random taste ($\{C\}$ or $\{V\}$). 
And similarly, looking at a candy having a specific taste will result in seeing a random color ($\{R\}$ or $\{G\}$).

Einstein believed that even if one general property must be randomized, its actually not --- its specific property is still \emph{there}, but well-hidden. 
Hence a \emph{deeper theory} (beyond quantum) could reveal both general properties with their explicit specific properties, so the candies eventually will follow (in the eyes of a futuristic scientist) classical intuition and eventually --- will be identical to Svozil's chocolate balls.

Today --- due to a theorem known as Bell's theorem~\cite{bell1964einstein}, and various confirming experiments (the first done by Aspect~\cite{Aspect1982,Aspect1982a}) --- the modern view is that we do not follow Einstein's belief anymore. 
As understood today, when Alice asks for a qandy with a definite taste, the machine would produce one with Alice's desired taste, but with NO DEFINED COLOR, and if Alice wants a qandy with a definite color, the machine would produce one with Alice's desired color, but with NO DEFINED TASTE! 
The complementarity rule is so deeply inherent in Jacobs' qandies model that we may say the other general property does not exist at all (it is not even random), and it only becomes random if an observation is made.


Finally, the qandy model illustrates that an observation alters the qandies' properties, leading to a no cloning principle and to secure QKD.

\subsection{Orthogonality, and No-Cloning}\label{Subsec:no-cloning}
We now discuss an important implication of these qandies, which will be useful in explaining the BB84 protocol. 
Here, the key concept is \emph{orthogonality}: we say that two qandy states are orthogonal to each other when they are different states of the same general property:
$\{R\}$ and $\{G\}$ are orthogonal states, and $\{C\}$ and $\{V\}$ are also orthogonal states.
Crucially, orthogonal states can be \emph{perfectly distinguished} if an observation is made to distinguish them. I.e., 
if Alice sends to Bob a color qandy, and tells bob it is a color qandy, then Bob can distinguish whether the qandy is red or green by looking at it. 
On the other hand, if Bob is given randomly a red or chocolate qandy, then no observation can determine, with 100\% chance of being correct, the specific property of that qandy.

The ability of identifying a particular state directly translates to the ability of copying (cloning) a qandy.
Thus, the existence of non-orthogonal states results in the impossibility to copy a random qandy that can be in arbitrary states.
This is named the no-cloning principle and is central to the safety of many quantum cryptography protocols, as we will now see.

\subsection{BB84 QKD with Qandies}\label{BB84-qandy}
We now demonstrate how Alice and Bob may utilize qandies to achieve the task of secure key distribution, using the BB84 protocol.
First, prior to sending anything, they determine a set of rules that assign binary values to states of each specific property; for example, they may assign 0 to red color and chocolate taste, and 1 to green color and vanilla taste.
This would allow their communication to result in a string of bits.
Then, when they are later separated and would like to exchange a key to encrypt their message, they repeatedly perform the following: Alice randomly presses one of the four buttons on her qandy machine, records her choice, and has the qandy delivered to Bob.
Upon receiving, Bob randomly decides what he would like to do with it: he can either ``look'' and record its color, or ``taste'' and record its taste.
He then translates the result into 0 or 1 and records the bit value \emph{as well as} his action.

After this has been repeated for a sufficient number of times, Alice and Bob will communicate through the classical channel to compare their preparation and observation \emph{methods} for every qandy\footnote{The classical channel is assumed to be insecure yet unjammable.}.
Specifically, Alice tells Bob whether she prepared a qandy with definite color or a definite taste, and Bob tells Alice whether he tasted or looked.
Note that they do not reveal their bit values to each other.
They keep the result if the two methods match, and discard otherwise\footnote{If Bob can keep the qandies un-measured in a qandies' memory, the bit rate can be doubled, as he will taste or look only after learning Alice's preparation method.}.
For example, if for qandy number 17 Alice generated a qandy with definite color (say $\{R\}$), and Bob measured a general property of a color, they will share an identical knowledge (unless natural noise or an eavesdropper destroyed it and input an error).
This establishes a string of random bits between the two parties. 

The safety of this protocol can be understood from the no-cloning principle.
Let us analyze a specific eavesdropping attack --- the ``measure-resend attack'':
Suppose Eve gambles and looks at qandy number 17. She will see $R$ and hence get full information. 
She then creates a new copy of Alice's red qandy and sends to Bob, and Bob observes red as he is supposed to.
However, we know that, in general, learning or cloning is impossible as discussed earlier; indeed, if Eve gambles badly and decides to taste qandy number 17, she will learn an irrelevant information (a random taste) and resend a definite taste, hence ``corrupt'' the original bit that Alice intended to send. With a probability of 50\% Bob will see $G$ instead of $R$, i.e., Eve generated noise.

From \cref{BB84-table} we see that overall, Eve's measure-resend attack will result in about $25\%$ mismatched bits among those shared between Alice and Bob, which shall be revealed to them by comparing a sufficiently long subset of their bits through the classical (insecure yet unjammable) channel.
In this case, Alice and Bob can discard their shared bits, and re-start this procedure again (via a different qandies channel) to obtain a new bit string, and verify (as we saw) that no eavesdropper is present. 
It has been proven that a secret key can be obtained via post-processing of the data, as long as the error-rate is smaller than around 10\%~\cite{Shor2000}.

\begin{table}[!ht]
	\centering
	\caption{Agreement probability between Alice and Bob due to the measure-resend attack. The table entries indicate actions taken by the eavesdropper as well as their occurring probabilities. A green cell implies agreement between Alice and Bob, whereas a red cell implies a disagreement. In 25\% of the cases, their bits do not agree.}	
	\begin{tabular}{ |c|c|c|c|c| } 
		\hline
		\backslashbox{Bob}{Alice} & \multicolumn{2}{|c|}{Color} & \multicolumn{2}{|c|}{Taste} \\ 
		\hhline{*{5}{-}}
		\multirow{2}{5em}{Look} &  \multicolumn{2}{|c|}{ \cellcolor[HTML]{9AFF99} Look, 50\% } & \multicolumn{2}{|c|}{\multirow{2}{*}{Discard}} \\
		\cline{2-3}
		\multicolumn{1}{|c|}{} & \cellcolor[HTML]{FFCCC9} Taste, 25\% & \cellcolor[HTML]{9AFF99} Taste, 25\% & \multicolumn{2}{|c|}{} \\
		\hline
		\multirow{2}{5em}{Taste} & \multicolumn{2}{|c|}{\multirow{2}{*}{Discard}} & \cellcolor[HTML]{FFCCC9} Look, 25\% & \cellcolor[HTML]{9AFF99} Look, 25\% \\ 
		\cline{4-5}
		& \multicolumn{2}{|c|}{} & \multicolumn{2}{|c|}{ \cellcolor[HTML]{9AFF99} Taste, 50\%} \\
		\hline
	\end{tabular}

	\label{BB84-table}
\end{table}

\subsection{A Few Direct Extensions}\label{Add-QKD-qandy}

Two simple variants of BB84 are presented to show that making the properties somewhat more classical is possible, while keeping the goal of secure communication. If qandies had only the color property and not the taste (or vice versa --- if had only the taste), a classical description would be immediate, cloning would be possible, and all parties, the sender the receiver and any potential eavesdropper will see everything --- the green or the red. 
One variant of BB84 that is somewhat close to be classical works when taste exists but qandies with taste are only rarely used. 
Such a protocol can still be proven secure~\cite{Lo2005}.
Another variant suggested the possibility that the tastes exist, but the machine used by the sender can only generate two colors (red and green) and just one taste (say, chocolate). 
As long as the receiver can detect both colors and both tastes, this protocol is secure~\cite{Mor1998}. 
This protocol had later led on to many protocols in which one party can only be aware of colors, and still --- such protocols (named semi-quantum key distribution) are secure.

A much more famous and well known protocol, the B92 protocol (when adjusted to qandies), uses just a single color and a single taste. 
In the B92 protocol~\cite{bennett1992quantum}, Alice only prepares two types of qandies: red and chocolate, which she and Bob agreed to stand for bits 0 and 1 respectively.
When Bob receives a qandy, he randomly chooses to taste or look, but does not reveal his choice to Alice as in BB84.
Instead, he publicly announces to Alice whether he \emph{succeeds} or \emph{fails} for every qandy, where the success condition is defined as \emph{tasting vanilla} or \emph{seeing green}.
He then associates seeing green with the bit 1 because Alice never prepared green and red could not lead to seeing green (only chocolate could); and he associates vanilla with 0 because Alice never prepared vanilla and chocolate could not lead to tasting vanilla (only red could).
After sending many qandies, as in BB84, they compare a subset of the bits to determine whether an eavesdropper is present, and final post-processing is used to generate a final key.

We can easily extend the qandies model to have more than two colors and tastes leading to additional key distribution protocols; some of those are actually ``beyond quantum'' in one sense or another.

For example, if instead of trying to imitate quantum bits--qubits, we try to imitate quantum trits--qutrits, we simply add one additional color (say, blue) and one additional taste (say peanut butter). Interestingly, if we follow restrictions imposed by quantum theory, we must have the same number of colors and tastes. 
Of course Alice and Bob are not forced to use all three colors and tastes in such a case, but once we define the candies to be analogous to qutrits, the eavesdropper can make use of that extended qandies space. 

Another fascinating extension is to go back to just two colors and two tastes, but add a third property, say texture, to yield ``the six-state protocol''\footnote{Interestingly having just two specific properties (two colors etc etc) and having four general properties --- e.g. color, taste, texture, and smell, is not consistent with the rules of quantum theory.}.
The six-state protocol is a generalization of the BB84 protocol, where 3 pairs of orthogonal states (each pair belonging to one general property) of qandies are used; this requires a slightly more general machine that is capable of preparing qandies with 3 general properties, a color, a taste, and a texture --- the texture of a qandy can be, for example, soft or crunchy.
As in BB84, the machine can only prepare a qandy with one definite specific property, and one can only learn about one general property of a qandy. 
And if a qandy is prepared with one specific property (say the the taste of vanilla) an attempt to look at it or to ``feel'' its texture must yield a random outcome, and the person eating it feels no softness or crunchiness.
The resulting QKD protocol is (in some sense) more secure than BB84: if we assume Eve applies an observation and sends according to her outcome, then among the bits intercepted by Eve, for which Alice and Bob used the same general property, $1/3$ of them would result in an inconsistency between Alice and Bob.
This is higher than in the BB84 protocol, meaning that there exists a level of natural error such that BB84 cannot be made secure while the six-state protocol can be made secure. 
Of course, this comes at some cost --- increasing the total number of qandies that Alice must share with Bob, due to the increased number (i.e. $2/3$) of discarded cases.


\section{Multi-qandies and Pseudo-Entanglement}\label{Sec:qandies-concepts}
We have demonstrated the principle of complementarity using the qandy picture, and illustrated why the BB84 protocol is safe.
We now exploit other quantum-inspired concepts that can be demonstrated by going far beyond Jacobs' qandy model. Specifically we shall consider here multi-qandy state, and qandies-correlations (i.e.  pseudo-entanglement), closely related to quantum entanglement.

It is easy to imagine a system of multiple qandies: for example, a system of a chocolate qandy and a green qandy can be described as $\{C\}_{1} \{G\}_{2}$, where the subscript denotes the qandy's number.
The two qandies live in different ``spaces'', denoted by the separation of the two states $\{\ \}_1\{\ \}_2$. 
As we will see, we can define new types of multi-qandy system that goes beyond a group of independent qandies.
In some sense, this demonstrates another complementarity principle, as we clarify by the end of this section.

\subsection{Correlated Qandies and Pseudo-entanglement}\label{Subsec:qandies-correlations}

Quantum entanglement is a phenomenon often called ``spooky action at a distance''. 
In its simplest form, it describes the behavior of two spatially separated systems where measurement outcomes on both systems are correlated in a weird (non-classical) way.
We can partially mimic it using two qandies.

To demonstrate this, we need to imagine a qandy-producing machine of a new type, which generates (in addition to the previously described single qandies) four types of correlated pairs of qandies. 
One type of correlated pairs, denoted as $\{\phi_+\}$, has the property that if the same test-action is performed on both qandies, the outcomes are always random but identical.
If both are looked at, then the outcome would always be either (red, red) or (green, green), each occurring with $50\%$ probability; similarly, if both are tasted, then the outcome would always be either (chocolate, chocolate) or (vanilla, vanilla), each occurring with $50\%$ probability.
On the other hand, if one qandy is tasted while the other is looked at, each outcome is random on its own, and there are no correlations between the two outcomes. 
These correlated qandies are different than the qandies of the earlier sections in the sense that each qandy in the pair can no longer be assigned a definite taste or color, and only the correlations are now defined.

We define the other correlated qandy pairs similarly as follows:
\begin{itemize}
    \item $\{\phi_+\}$ as above; 
    \item $\{\psi_+\}$: If both candies are looked at, they present opposite colors, but whenever they are tasted, they present identical tastes;
    \item $\{\phi_-\}$: If both candies are looked at, they present identical colors, but whenever they are tasted, they present opposite tastes;
    \item $\{\psi_-\}$: If both candies are looked at, they present opposite colors, and whenever they are tasted, they present opposite tastes.
\end{itemize}

As before, in all these cases, if one qandy is tasted while the other is looked at, there are no correlations between the outcomes, and the outcomes are totally random.
These qandies possess what we call pseudo-entanglement.

We can easily mimic a variant of BB84 using such a machine. 
This, in its quantum version, is called the EPR scheme~\cite{Ekert1991quantum,bbm92}. 
Here, Alice prepares a correlated pair $\{\phi_+\}$. 
She then sends one qandy to Bob and keeps the other one. 
Each of them chooses whether to taste the qandy or look at it. 
They expect the same result if both performed the same action, and they expect random and uncorrelated results if they performed different actions. 
After repeating a sufficient number of times, they compare their actions and keep only the results from the cases which they performed the same action. 

In all the cases where Alice and Bob performed the same action, their result is identical, random,
and secure against Eve. This ensures the safety of the protocol. Importantly, even if Eve prepares the qandy pairs (instead of Alice), she does not know what answer Alice and Bob would get.

\subsection{Quantum Bit Commitment}\label{Sec:bit-commitment}

With correlated qandies, we can show another interesting quantum cryptographic protocol; unfortunately, the protocol is actually insecure, and the simplest and most typical attack showing its insecurity can be fully described using pseudo-entangled qandies.
We first consider a goal called bit commitment, where Alice would like to first deliver a bit to Bob, and reveal the bit at a later time.
However, the two parties may not trust each other: for example, Bob may suspect that Alice could change her bit after it has been sent, and Alice may suspect that Bob could somehow know the bit before she decides to reveal it. We divide the protocol into a commit stage and a reveal stage.

One quantum scheme (along with a proof of its insecurity!) was proposed by Bennett and Brassard in 1984~\cite{bennett1984quantum}. 
Their scheme works as follows: In the commit stage, Alice prepares $n$ qandies with definite colors (each randomly chosen to be red or green) if she wants to commit a bit 0, or $n$ qandies with a definite taste (randomly chocolate or vanilla) if she wants to commit 1.
She keeps a record of what she has prepared (a list of red, chocolate, green, etc., for example), and delivers the $n$ qandies to Bob, without telling him how she had prepared them.

At the reveal stage, she tells Bob her choice of preparation of all $n$ qandies, which effectively uncovers the bit.
Bob, who is assumed to have a memory where he keeps the qandies\footnote{There is a similar protocol if Bob has no memory, with a slightly different analysis.}, can verify that she did not cheat (i.e., changed the bit she originally intended to send) by asking Alice what he should get for each qandy.

The safety consideration of the protocol is as follows.
Clearly, if Bob randomly looks at or tastes the qandies, the result is completely random and meaningless, as the information is fully encoded in Alice's preparation method. Thus, the best strategy for Bob is to do nothing at the commit stage, and then he can fully verify Alice's opening at the reveal stage.
So Alice can be sure that Bob does not have any information on her bit.
On the other hand, since Alice cannot change the qandies once they're in Bob's hand, if Alice did cheat by telling Bob the opposite method, she would need to guess Bob's random results, which has a success probability of $2^{-n}$.
Therefore, cheating can be prevented with arbitrarily high probability if Bob requires Alice to verify a longer and longer bit string.

However, the above argument does not hold if Alice is capable of preparing pseudo-entangled qandies.
Alice can cheat by doing the following --- she prepares $n$ pairs of qandies in $\{\phi_{+}\}$, sends one of each pair to Bob and keeps the other one until she wants to reveal.
At the reveal stage, she can tell Bob any bit she ``was commiting'' (either color or taste), and when Bob would like to verify, she can taste or look at (according to her choice at the last minute) every qandy, and tell Bob the results to verify on the other member of each pseudo-entangled pair.
Because of the perfect correlation in both color and taste for $\{\phi_{+}\}$, Bob would always be convinced, while Alice has the complete freedom to cheat and decide what she wants to send long after the committing stage.
Therefore, the protocol is no longer safe when pseudo-entangled qandies are available to Alice; a similar insecurity argument is true in the quantum scenario, and is even generalized to any possible quantum bit-commitment protocol~\cite{mayers1997unconditionally}.

This protocol and the attack demonstrate nicely the spooky action at a distance! 
Alice controls Bob's qandies, exactly to the desired level, even though she has no access to them at the stage when she decides on how to control them.

\section{Conclusion} \label{Sec:Discussion}
In this work, we have introduced and extended the quantum candy model proposed by Jacobs, and used it to demonstrate some of the most important concepts in quantum information science and quantum cryptography in an approachable manner.
Our work is meant to be a useful tool for introducing quantum science to the general public.
We emphasize that the qandy model does not always need to follow quantum theory and can be extended in many ways. 
We leave this to the full paper, as well as the well established connection to quantum theory, which mainly contains quantum superposition and unitary transformations. 

\section*{Acknowledgments}J.L. and T.M. thank the Schwartz/Reisman Foundation. J.L. is supported by NSERC Canada. T.M. was also partially supported by Israeli MOD.

%
%
\bibliographystyle{splncs04}
\bibliography{candyQKD}

\end{document}